\documentstyle[12pt]{article}
\def\eea{\end{eqnarray}}
\def\bea{\begin{eqnarray}}
\def\eeas{\end{eqnarray*}}
\def\beas{\begin{eqnarray*}}
\def\ee{\end{equation}}
\def\be{\begin{equation}}
\def\eeq{\end{equation}}
\def\beq{\begin{equation}}
\def\eaq{\end{equation}}
\def\baq{\begin{equation}}

\def\fpi2{\mbox{F$_\pi$}^2}

\def\mpi2{{m_\pi}^2}
\def\mk{m_K}
\def\mk2{{m_K}^2}

\def\fk2{\mbox{F$_K$}^2}

\renewcommand{\thefootnote}{\fnsymbol{footnote}}
\begin{document}
\begin{titlepage}

\begin{center}
\hfill JLAB-THY-98-25 \\
\hfill LA PLATA-TH 98/11 \\
\hfill {\it nucl-th/9806048} \\
\vspace*{1.1cm}

{\large\bf EQUATION OF STATE OF HADRONIC MATTER WITH
DIBARYONS IN AN EFFECTIVE QUARK MODEL}
\vskip 1.1cm

{Ricardo M. AGUIRRE $^a$\footnote{Electronic address:
aguirre@venus.fisica.unlp.edu.ar} and
Martin SCHVELLINGER$^{a,b,c}$\footnote{On leave from
University of La Plata. \\
Electronic addresses: schvell@cebaf.gov  and  martin@venus.fisica.unlp.edu.ar
}}
\vskip .2cm
{\it
$^a$ Physics Department, University of La Plata, C.C. 67, (1900)
La Plata, Argentina. \\
$^b$ The Nuclear/High Energy Physics Research Center, Hampton University, \\
Hampton, VA 23668, USA.\\
$^c$ Thomas Jefferson National Accelerator Facility, 12000 Jefferson Avenue,
Newport News, VA 23606, USA. \\
}

\vskip 1. cm
{\bf ABSTRACT}\\
\begin{quotation}

The equation of state of symmetric nuclear matter with the
inclusion of non-strange dibaryons is studied. We pay special
attention to the existence of a dibaryon condensate at zero
temperature. These calculations have been performed in an
extended quark-meson coupling model with density-dependent
parameters, which takes into account the finite size of
nucleons and dibaryons. A first-order phase-transition
to pure dibaryon matter has been found. The corresponding
critical density is strongly dependent on the value
of the dibaryon mass. The density behavior of the nucleon
and dibaryon effective masses and confining volumes have
also been discussed.

\end{quotation}
\end{center}



\end{titlepage}
\renewcommand{\thefootnote}{\arabic{footnote}}
\setcounter{footnote}{0}

Nuclear matter properties have been modeled by several kind of
local relativistic effective lagrangians which basically use
point-like representations of nucleons and mesons  as the
relevant degrees of freedom. The differences are essentially
coming from the nucleon-scalar meson interaction. It should be
desirable to describe nuclear matter from a fundamental
theory like QCD, but it is wellkown that in the low energy limit
QCD becomes non-perturvative. This situation has motivated the
development of several effective models of quark interaction,
among them there is the MIT bag model \cite{CH74}. This model
predicts the existence of some multibaryons and strange exotics
\cite{FA84}. The early work of Jaffe \cite{JA77} has concentrated
the attention of theoretical studies \cite{FA84,GO98}
and the further development of experimental research looking for
signals of strange and non-strange dibaryons.
In reference \cite{JA77} it has been shown that in the scheme of the
MIT bag model the gluon-exchange force should be responsible of the
existence of a stable six-quarks bound state. This particle,
the so-called $H$-particle, is a flavor singlet $J^\pi = 0^+$ dihyperon
with a mass of $2150 MeV$ and strangeness $-2$.
The problem has also been treated in the non-perturbative framework
of the Skyrme model where dibaryons have been considered as axially
symmetric skyrmions \cite{SC94}. Non-strange as well as strange dibaryons
and another multibaryons have been studied as multiskyrmions \cite{SC98}.
The experimental activity concerned to this search has been increased in
the last years, recent experiments have been developed at TRIUMF and CELCIUS
\cite{EXP}. Non-strange dibaryons which have a small width have been described
as promising candidates for experimental searches \cite{FA971}.

On the other hand, models based on the quantum field theory of hadrons
including non-strange dibaryons as effective degrees of freedom have been
extensively studied \cite{FA97}, obtaining very interesting effects on the
binding energy per particle as well as on the equation of state (EOS) of
the system. In these hadronic effective lagrangians  nucleons and dibaryons
are treated as point-like particles represented by two independent
effective fields, both of them interact by the exchange of scalar and vector
neutral mesons using two different sets of coupling constants.

The purpose of this work is to include finite size effects on the EOS,
taking into account the quark structure of the particles. As we will
show later, several features of our results
are in agreement with those obtained in reference \cite{FA97}.
Therefore, this fact seems to support that the dibaryon condensate
is essentially a model-independent issue. We have selected the so-called
quark-meson coupling model (QMC) \cite{GU88},
which in some way satisfies the above mentioned requirements.
In this scheme we can perform a simultaneous description of nucleons
and dibaryons using only current quarks and mesons as effective
degrees of freedom. The model was early developed by Guichon \cite{GU88}
and it has been extensively applied to calculate nuclear matter
\cite{SA94} as well as finite nuclei \cite{SA96} properties with successful
results. It has also
been used to evaluate nucleon structure functions \cite{MIC92}. The naturalness
of the QMC model has been studied using the dimensional analysis \cite{TH97}.
Recently the density dependence of the parameters of the MIT bag 
have been phenomenologically modeled \cite{JI96}, alternatively a
 relationship with observables evaluated in
the quantum field theory 
of hadrons have been stablished \cite{AG97,AG97EMC}.

In this work we have developed an extension of the QMC model described
in reference \cite{AG97}, by including dibaryons represented as spherical
MIT bags confining six quarks.
The extended QMC lagrangian density with quark fields $q_{\alpha}(x)$
coupled to scalar $\sigma(x)$ and vector $\omega_{\mu}(x)$ neutral
mesons, is written as follows
\begin{eqnarray}
&{\cal{L}}_{QMC}(x) = {\cal{L}}_{N}(x) + {\cal{L}}_{D}(x)
+ {\cal{L}}^0_{Mesons}(x) \ , & \\
&{\cal{L}}_{N}(x) = \left( {\cal{L}}^{0}_{N}(x) - B_1 \right) \Theta_{VN}
- \frac{1}{2} \sum^3_{\alpha=1}{\bar{q}}_\alpha(x) q_\alpha(x)
\Delta_{SN} \ , & \\
&{\cal{L}}_{D}(x) = \left( {\cal{L}}^{0}_{D}(x) - B_2 \right) \Theta_{VD}
- \frac{1}{2} \sum^6_{\alpha=1}{\bar{q}}_\alpha(x) q_\alpha(x)
\Delta_{SD} \ . &
\label{LAGR1}
\end{eqnarray}
Here $\Theta_{VN}$ and $\Theta_{VD}$ are the radial non-overlapping
step functions which schematically confine the quarks inside
spherical bags for nucleons and dibaryons, respectively.
$B_1$ and $B_2$ are the so-called MIT bag constants
associated with these particles.
Within the standard QMC treatment $B$ is a constant,
however it can be considered as function of the
baryonic density, $\rho_B$. The terms proportional to the surface
delta functions $\Delta_{SN}$ and $\Delta_{SD}$
ensure a zero flux of quark current through the bag surface. In this
lagrangian  we have defined the following terms
\begin{eqnarray}
&{\cal{L}}^{0}_{N}(x) = \sum^3_{\alpha=1} {\bar{q}}_\alpha(x)
\left( i \gamma^\mu{\partial}_\mu - m_\alpha + g_\sigma \sigma(x) -
g_\omega \gamma^\mu \omega_\mu(x) \right) q_\alpha(x) \ ,& \\
&{\cal{L}}^{0}_{D}(x) = \sum^6_{\alpha=1} {\bar{q}}_\alpha(x)
\left( i \gamma^\mu{\partial}_\mu - m_\alpha + g_\sigma \sigma(x) -
g_\omega \gamma^\mu \omega_\mu(x) \right) q_\alpha(x) \ ,& \\
&{\cal{L}}^0_{Mesons}(x) =  \frac{1}{2} [ \partial^\mu \sigma(x)
\partial_\mu \sigma(x) - m_\sigma^2 \sigma^2(x) ]
- \frac{1}{4} F^{\mu \nu}(x) F_{\mu \nu}(x) +
\frac{1}{2} m_\omega^2 \omega^\mu(x) \omega_\mu(x) \ .&
\label{LAGR2}
\end{eqnarray}
Here $g_{\sigma}$ and $g_{\omega}$ are the quark-meson
coupling constants associated with $\sigma(x)$ and $\omega_{\mu}(x)$,
respectively.
In order to get the minimal set of free parameters we have assumed
that quarks do not distinguish baryon or dibaryon bags, {\em i.e.} the
coupling constants are the same in both of the equations for baryons and
dibaryons. In that follows we deal with $u$ and $d$ massless
quarks, furthermore we introduce the index $\nu = 1$, $2$ to
label quantities related to nucleons and dibaryons, respectively.

The normalized quark wave function for the fundamental state in
a spherical bag of radius $R_{\nu}$ is given by
\beq
q_{\nu}({\vec{r}},t) = {\cal{N}}_{\nu}
e^{-i {\epsilon}_{\nu} t/R_{\nu}}
\times \left( \begin{array}{c} {j_0(y_{\nu} r/R_{\nu})} \\
i {\beta}_{\nu}
{\vec{\sigma}} \cdot $\^{r}$ j_1(y_{\nu} r/R_{\nu})  \end{array} \right)
{{{\chi}_\nu}\over{\sqrt{4 \pi}}} \ ,
\label{QUARKWF}
\eeq
where $r$ is the distance from the center of the bag,
${\chi}_\nu$ is the quark spinor and the normalization constant is
\beq
{\cal{N}}_{\nu} = \frac{y_{\nu}}{\sqrt{2 R^3_{\nu} j^2_0(y_{\nu}) [\Omega_{\nu}
(\Omega_{\nu}-1) + R_{\nu}
m^*_\alpha/2]}} \ .
\eeq
The parameter associated with the quark mass is
$m^*_\alpha = m_\alpha -g_{\sigma} {\bar{\sigma}}$, and the
 energy eigenvalue is written as
\beq
{\epsilon}_\nu = \Omega_{\nu} + g_{\omega} \nu {\bar{\omega}} R_{\nu} \ ,
\eeq
where ${\Omega_{\nu}} = \sqrt{y_{\nu}^2 + (R_{\nu} m^*_\alpha)^2}$, and
$\bar{\sigma}, \bar{\omega}$ are the mean values of meson fields
calculated in the Mean Field Approximation.
The $y_{\nu}$ variable is fixed by the boundary condition at the bag
surface $j_0(y_{\nu}) = {\beta}_\nu j_1(y_{\nu})$ as in reference \cite{CH74}
and ${\beta}_\nu = \sqrt{({\Omega_{\nu}} - R_{\nu} m^*_\alpha)/
({\Omega_{\nu}} + R_{\nu} m^*_\alpha)}$. The mass associated with the bag
is given by
\begin{equation}
M_{\nu} = \frac{3 \nu \Omega_{\nu} - z_{0 \nu}}{R_{\nu}} +
\frac{4}{3} \pi B_{\nu} R^3_{\nu} .
\label{MASS}
\end{equation}
The $B_{\nu}$ are the bag constants already introduced in Eqs.(2) and (3),
while $z_{0 \nu}$ takes into account the zero point energy of the bag.

The usual procedure in QMC is to fix the nucleon bag
parameters at zero baryon density to reproduce the experimental nucleon
mass $M_1 = M_N = 939 MeV$, simultaneously it is required that the equilibrium
condition  $d M_1({\bar{\sigma}})/dR=0$ must be fulfilled.
An analogous method should be applied to dibaryons, however the
experimental value of in-vacuum dibaryon mass $M_D$ has not
been definitely confirmed yet, and at the present, only theoretical estimates
are availables. Therefore, we leave $M_D$ as a parameter of the model.
Since the $B_{\nu}$ are related to vacuum properties we assume that
$B_1 = B_2 = B$, at all densities. Although the paramaters $z_{0 \nu}$
could also have a density dependence, in a previous work \cite{AG97}
it was found that $z_{0 1}$ remains approximately constant at the baryon
densities below four times the nuclear matter saturation
density. Consecuently we have taken
$z_{0 \nu}$ as constants fixed at zero density for each kind of bag.
Under these assumptions one can immediately get a relation for
masses and radii : $R_2 = (M_2/M_1)^{1/3} R_1$.

In order to obtain the density dependence of $B$ we have stablished an explicit
relationship between nuclear matter observables evaluated in the extended QMC
model and in pure hadronic models, as it has been described in detail in
\cite{AG97}. In the present work we have selected the Zimanyi-Moszkowski model
\cite{ZM} to describe the hadronic sector (as in reference \cite{AG97}).
Considering that the effective nucleon mass predicted by the extended QMC model
and the hadronic model must be the same, together with the fact that the outward
quark-momentum on the bag surface must be compensated by the hadronic momentum
going inside, one gets the following equations
\beq
M_1 = M^* \ ,
\label{EQUALMASS}
\eeq
\beq
P_{bag} = P_{had} \ ,
\label{EQUALPRESS}
\eeq
which are valids at each value of baryon density. Here $M^*$ and $P_{had}$ are
the effective nucleon mass and pressure of nuclear matter in the hadronic model,
while
$P_{bag} = - (dM_1/dR_1)/4 \pi R_1^2$ is the pressure in the nucleon bag.
Using these equations and $m_{\sigma}=550 MeV$,
$m_{\omega}=783 MeV$, $g_{\sigma}=4.576$,
$g_{\omega}=2.222$ and $R_1^0=0.8 fm$ we have obtained the density
behaviour of $B=B(\rho_B)$ shown in Fig. \ref{BRFig}.

Once we have dinamically derived the parameter $B$ as a function of
the baryon density, we can perform the calculations for nuclear matter
with dibaryons. Firstly, we evaluate the energy per baryon of
a system composed by symmetric nuclear matter with density $\rho_N$, and
dibaryons with density $\rho_D$. Since each dibaryon carries baryon number
two, the baryon density must be $\rho_B=\rho_N+2 \rho_D$.
Using the quark wave functions of Eq.(\ref{QUARKWF}) one can construct the
antisymmetrized (symmetrized) nucleon (dibaryon) physical states and
evaluate the expectation value of the energy density $\cal{H}$ derived from
the lagrangian of Eq.(1). For uniform matter the energy per baryon is given by
\beq
\epsilon   = {\cal H}/\rho_B  =\frac{1}{\rho_B} \left[ \frac{ M_1^4}{ \pi^2}
F(\eta)+
\frac{m_{\sigma}^2 \bar{\sigma}^2}{2}+
\frac{m_{\omega}^2 \bar{\omega}^2}{2} + M_2 \rho_D \right] \ ,
\eeq
where $\eta = k_F/M_1$, $k_F$ is the Fermi momentum for the nucleons and it is
related to the nucleon density by $\rho_N =2 k_F^3/ 3 \pi^2$, while
$F(\eta)= \eta \sqrt{\eta^2 + 1} (2 \eta^2 + 1) -
Log (\eta + \sqrt{\eta^2 + 1})$. Chemical potentials for nucleons and
dibaryons are given by
$\mu_N = \sqrt{k_F^2 + M_1^2} + 3 g_{\omega} \bar{\omega}$ and
$\mu_D = M_2 + 6 g_{\omega} \bar{\omega}$, respectively. Although the
dibaryons are bosons, the corresponding particle number is conserved
because they carry baryon charge and since in our lagrangian  we have not
included any decay mechanism. The pressure of hadronic matter at zero
temperature can be written as
\beq
P_{had} = \mu_N \rho_N + \mu_D \rho_D - \epsilon \rho_B \ .
\label{PRESSURE}
\eeq
The mean values of the meson fields have been evaluated by minimizing the
energy, {\em i.e.}
$\partial \epsilon/ \partial \bar{\sigma} = 0$ and $\partial \epsilon/ \partial
\bar{\omega} = 0$, obtaining
\beq
\bar{\sigma}=- \frac{1}{m_s^2} \left\{ 2 \frac{M_1^2}{\pi^2}\left[
4 M_1 F(\eta) - p_F \frac{dF}{d \eta}(\eta) \right]
 \frac{dM_1}{d \sigma}
+ \rho_D \frac{dM_2}{d \sigma} \right\} \ ,
\label{SIGMA}
\eeq
together with $\bar{\omega}=3 g_{\omega} \rho_B/m_{\omega}^2$.

In order to obtain numerical results we have arbitrarily selected the
value of the in-vacuum dibaryon mass $M_D = 1970 MeV$. We have used the meson
masses, free nucleon bag radius and the previously described procedure to
calculate several in-vacuum quantities such as the
parameters $z_{01}=3.27$, $z_{02}=6.27$, as well as the dibaryon bag radius
$R^{(0)}_2 = 0.954$ $fm^{-1}$, the usual boundary condition eigenvalue
$y^{(0)}_\nu = 2.042$ and the bag parameter $B^{(0)} = 0.5546$ $fm^{-4}$.
When the constants $z_{0 1}, z_{02}$ have been fixed and the density
dependent parameter $B$ has been obtained, we must find the effective bag radii
$R_1(\rho_B)$ and $R_2(\rho_B)$. To do that we have impossed the equilibrium
condition for bags immersed in hadronic medium in similar way as
Eq.(\ref{EQUALPRESS}), but now using Eq.(\ref{PRESSURE}) in the right hand side.
After a little algebra it gives
\beq
4 \pi R_{\nu}^4 P_{had} = - 3 M_{\nu} R_{\nu} + 4 (3 \Omega_{\nu} -
z_{0 \nu}) - 3 a_{\nu} \frac{2 a_{\nu} (\Omega_{\nu} - 1) + \Omega_{\nu}}
{2 \Omega_{\nu} (\Omega_{\nu} - 1) + a_{\nu}} \ ,
\label{EQUILIBRIUM}
\eeq
where we have defined $a_{\nu} = m^*_\alpha R_{\nu}(\rho_B)$. It must be noticed
that this condition becomes the standard equilibrium requirement of QMC model
for vanishing $\rho_B$.

Solving the boundary condition at the bag surface simultaneously with
Eqs.(\ref{SIGMA}) and (\ref{EQUILIBRIUM}),
for a given baryon density and several values of the dibaryon abundance
$Q = \rho_D / \rho_B$ we have evaluated the bag radii, masses, energy
density and pressure as functions of $\rho_B$ and $Q$.
The coupling constants	have been fixed
in order to reproduce at $Q = 0$ the saturation density
$\rho_0 = 0.15$ $fm^{-3}$ and the binding energy per particle
$\epsilon_0 = 16 MeV$, obtaining
$g_{\sigma}=4.6876$ and $ g_{\omega}=2.2967$. In Fig. 1
the results for the nucleon bag radius $R_1$ are shown. From this we can see
that at low densities the radius is an incresing function of
$\rho_B$ and it becomes a smoothly decreasing function at densities
around the saturation nuclear density. Furthermore, at densities below
$1.4 \rho_0$, $R_1$ is higher than in-vacuum radius,
notwithstanding for $Q = 0$ the relative increment of $R_1$ remains under the
value $2 \%$, as it is expected from y-scaling arguments in the analysis of
quasi-elastic electron scattering \cite{SICK}. Therefore, the use of
this radius and the nucleon effective mass $M_1$ is equivalent to have a
bigger radius and the
free nucleon mass, as it has been proposed in the reference of Sick \cite{SICK}.
Density variation of the radius is less
pronounced as $Q$ increases. A similar behaviour for the dibaryon bag radius
has been found.

To investigate the existence of a phase-transition from symmetric
nuclear matter to a pure dibaryonic state we have used the Gibbs criterion.
If there is a dynamical mechanism combining two
nucleons to give one dibaryon, at the phase-transition point the conditions
$2 \mu_N = \mu_D$ and $P_{nuc} = P_{dib}$ must be fulfilled.
Although we have not included such a reaction channel in our model,
we are concerned with the stable initial and final phases rather than with
any particular mechanism of dibaryon formation.
Here $P_{nuc}$ and $P_{dib}$ denote the pressure in pure nuclear and
pure dibaryon matter, respectively.
In Fig. \ref{POTQFig} the quantities $2 \mu_N$ for $Q=0$ and $\mu_D$
for $Q=0.5$ are represented as functions of the corresponding
pressures. The intersection point of these curves indicates that
phase-transition occurs at a critical pressure nearly
$P_0=8.8$ $10^{-2} fm^{-4}$.
At low pressure the local minimum of the Gibbs potential per particle
corresponds to pure nuclear matter, while at pressures beyond
$P_0$ the stable state corresponds to pure dibaryon matter.
In order to construct the EOS we have drawn Fig. \ref{PHASTFig} with the
pressure of Eq.(\ref{PRESSURE}), corresponding to $Q=0$ and $Q=0.5$ as
functions of baryon density. The value $P_0$ is reached at baryon densities
of $\rho_I=2.55 \rho_0$ in nuclear matter and
$\rho_{II}=3.18 \rho_0$ in the dibaryon condensate.
Therefore it is a first-order phase-transition
with a discontinuity jumping $0.63 \rho_0$ in the baryon density.
The horizontal segment $P=P_0$
between $\rho_I$ and $\rho_{II}$ represents two-phase coexistence.
The effect of the phase-transition is to reduce
the compressibility at high densities.
For example, the quotient of the thermodinamical compressibility
evaluated in the dibaryon condensate to its value in the nuclear matter
gives at $\rho_B=3.5 \rho_0$  the value $0.706$. This fact could be 
favorable for the collapse of very
massive stars \cite{COOPERS}. In Table I a comparison of the quotient of
thermodinamical compressibility at $Q = 0.$ and $Q = 0.5$ for several values
of $M_D$ at the transition point is shown.

In addition, we have studied how much is modified the phase-transition point as
the in-vacuum dibaryon mass $M_D$ is changed.
It has been found that the density $\rho_I$ is an increasing function of
the mass $M_D$, as it is shown in Table I. We must take into account
that dibaryons	have  not been observed at densities around the normal
saturation density $\rho_0$ and at sufficiently high densities our
model can not describe the quark-gluon plasma phase-transition. Therefore we
limit the search for the nuclear-dibaryon phase-transition at densities ranged
from $1.5 \rho_0$ to $4 \rho_0$. This requirement constrains the variation of
dibaryon mass to $1940 MeV < M_D < 2000 MeV$.

A comparison with the hadronic field the\-o\-ret\-i\-cal model of
re\-fe\-ren\-ce
\cite{FA97} shows that in the Hartree approximation a nuclear-dibaryon
heterophase appears rather than a pure condensate.
The Bose-condensate for $d'$ dibaryon appears at 3 times the saturation density
of nuclear matter. Independent calculations using the
quantum field theory of hadrons and dibaryons have been developed in
\cite{GL98}.
In this framework the transition density for nuclear and neutron matter
is $\rho_D/\rho_0=2.87$ and
$\rho_D/\rho_0=2.16$ respectively \cite{HORVATH}, similar results are found
in reference \cite{KA92}:
$\rho_D/\rho_0=2.69$ and $\rho_D/\rho_0=2.57$
for nuclear and neutron matter, respectively. Therefore our prediction
of $\rho_D/\rho_0=2.55$ is consistent with these values.

In this work we have studied the properties of a system composed of
symmetric infinite nuclear matter with dibaryons. This has been performed in a
theoretical framework which takes into account the finite size as well as
the quark structure of nucleons and dibaryons. The variation of particle
properties with baryon density have also been considered.
The nucleon mass shows a monotonic decreasing behavior for all the
densities and dibaryon abundances $Q$ studied here. At $Q=0$ nucleon
swelling is observed at densities below the normal nuclear matter density.
The bag radius increment is less than $2 \%$, in accordance with
theoretical estimates based on y-scaling interpretation of quasi-elastic
electron scattering \cite{SICK}.
A nuclear-dibaryon matter phase-transition is found at zero temperature.
The dibaryon condensate predicted in this work could be an intermediate
state before the transformation to quark matter.
If a dibaryon condensation takes place for baryon densities in the range
$1.5 < \rho_B / \rho_0 < 4$, then values $1940 MeV < M_D < 2000 MeV$
are predicted for non-strange dibaryon masses.
As a consequence of phase transition the compressibility of the
system is considerably reduced at high densities, as it should be
expected from astrophysical scenarios.

Extensions of the present work in order to describe strange dibaryons can be
inmediately
performed by inclusion of the color electric and color magnetic
interactions in the MIT bag model, also multiquark bags describing some exotic
states
of multibaryons could be treated in this framework.

This work has been supported in part by the grant PMT-PICT0079 of ANPCYT, Argentina.
M.S. has been supported in part by The Nuclear High Energy Physics Research Center of
Hampton University, USA, and by a grant of the
Fundaci\'on Antorchas, Argentina. We have been beneficted from stimulating 
discussions with J.E. Horvath
(IAG-USP, Brazil). M.S. is greatly indebted to J.L. Goity
(Jefferson Lab-Hampton University) for kind hospitality at Jefferson Lab.

\hfill
\eject

\newpage
\begin{table}
\begin{center}
\begin{tabular}{ccccc}
\hline \\
$M_D$ [GeV] & $\rho_I [fm^{-3}]$ & $\rho_{II} [fm^{-3}]$ &
$\kappa_{II} / \kappa_{I} $	\\

\hline
\hline

1.94 & 0.228 & 0.335  & 1.30  \\
1.95 & 0.283 & 0.380  & 1.09  \\
1.96 & 0.335 & 0.429  & 1.01  \\
1.97 & 0.383 & 0.477  & 0.98  \\
1.98 & 0.427 & 0.522  & 0.96  \\
1.99 & 0.470 & 0.566  & 0.94  \\
2.00 & 0.505 & 0.603  & 0.93  \\

\hline
\end{tabular}
\end{center}
\caption{ The baryon density $\rho$ and the quotient of thermodinamical
compressibility $\kappa= \partial P/\partial \rho_B$ evaluated at the
transition point as functions of the in-vacuum dibaryon mass $M_D$. The label
I (II) corresponds to pure nuclear (dibaryon) matter state.}
\end{table}

\newpage

\begin{figure}
\caption{
$B$ (dashed line) and the nucleon bag radius, $R_1$, for the
values $Q=0$, $0.25$, $0.33$, $0.47$ and $0.5$ (solid lines)
as functions of the baryon density, $\rho_B$.
The arrow indicates the increasing Q values. $R_N$ is the nucleon radius
at zero baryon density.}
\label{BRFig}
\end{figure}

\begin{figure}
\caption{
The dibaryon chemical potential (dashed line) evaluated at $Q=0.5$
and two times the nucleon chemical potential (solid line) at $Q=0$
plotted as functions of the total pressure of the system.
The phase-transition for in-vacuum dibaryon mass $M_D=1970$ MeV
occurs at a baryon density of $\rho_I=2.55$ fm$^{-3}$ in the nuclear
matter phase, corresponding to a baryon density of $\rho_{II}=3.17$
fm$^{-3}$ in the pure dibaryon phase.
}
\label{POTQFig}
\end{figure}

\begin{figure}
\caption{
Pressure as a function of baryon density for the dibaryon
abundances $Q=0$, $0.25$, $0.33$, $0.47$ and $0.5$ (dashed lines).
The arrow shows the increasing Q values. Solid line represents the
physical states of the system, corresponding to nuclear matter ($Q=0$)
below the relative density $\rho_B/\rho_0=2.55$
and to pure dibaryon matter ($Q=0.5$) above the value
$\rho_B/\rho_0=3.17$. The horizontal segment between these densities
represents a coexistence region between the two phases.
}
\label{PHASTFig}
\end{figure}

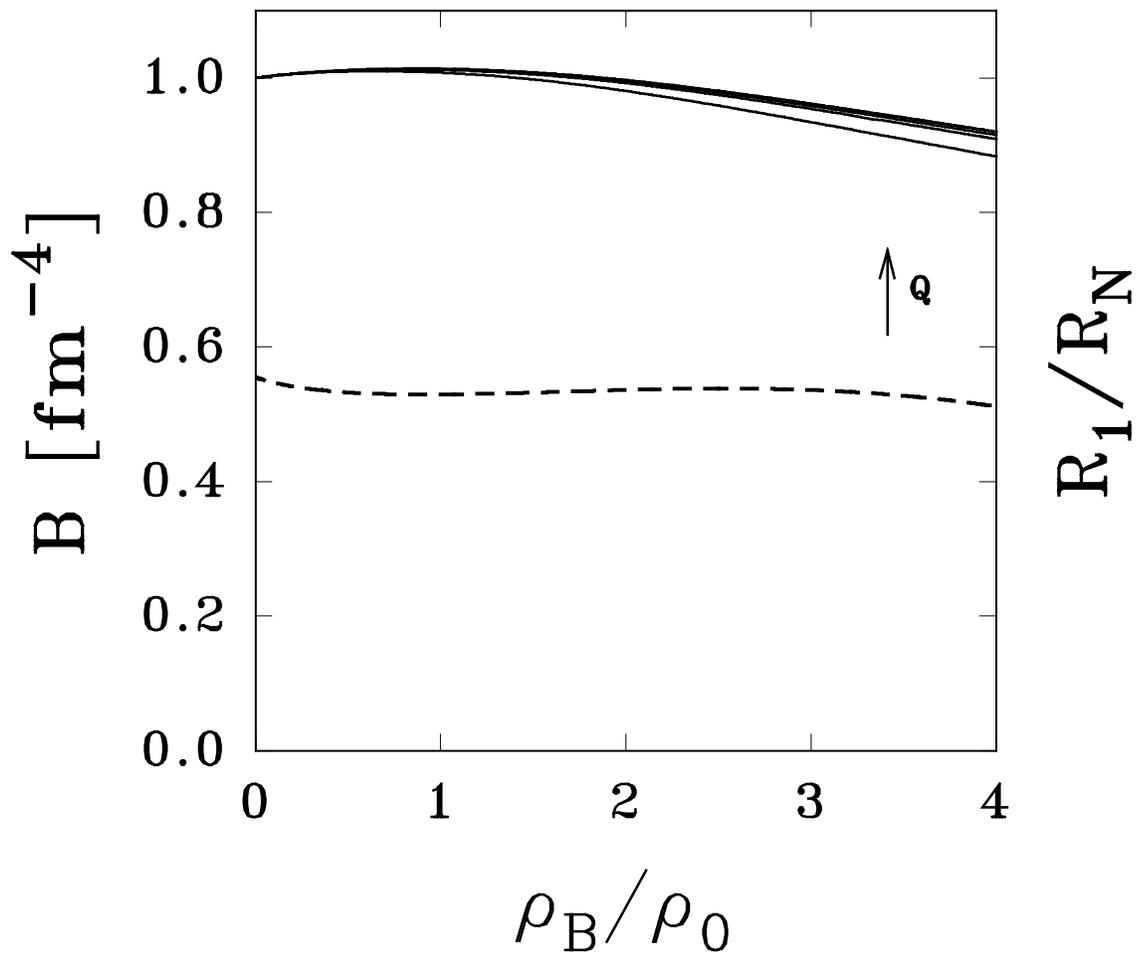
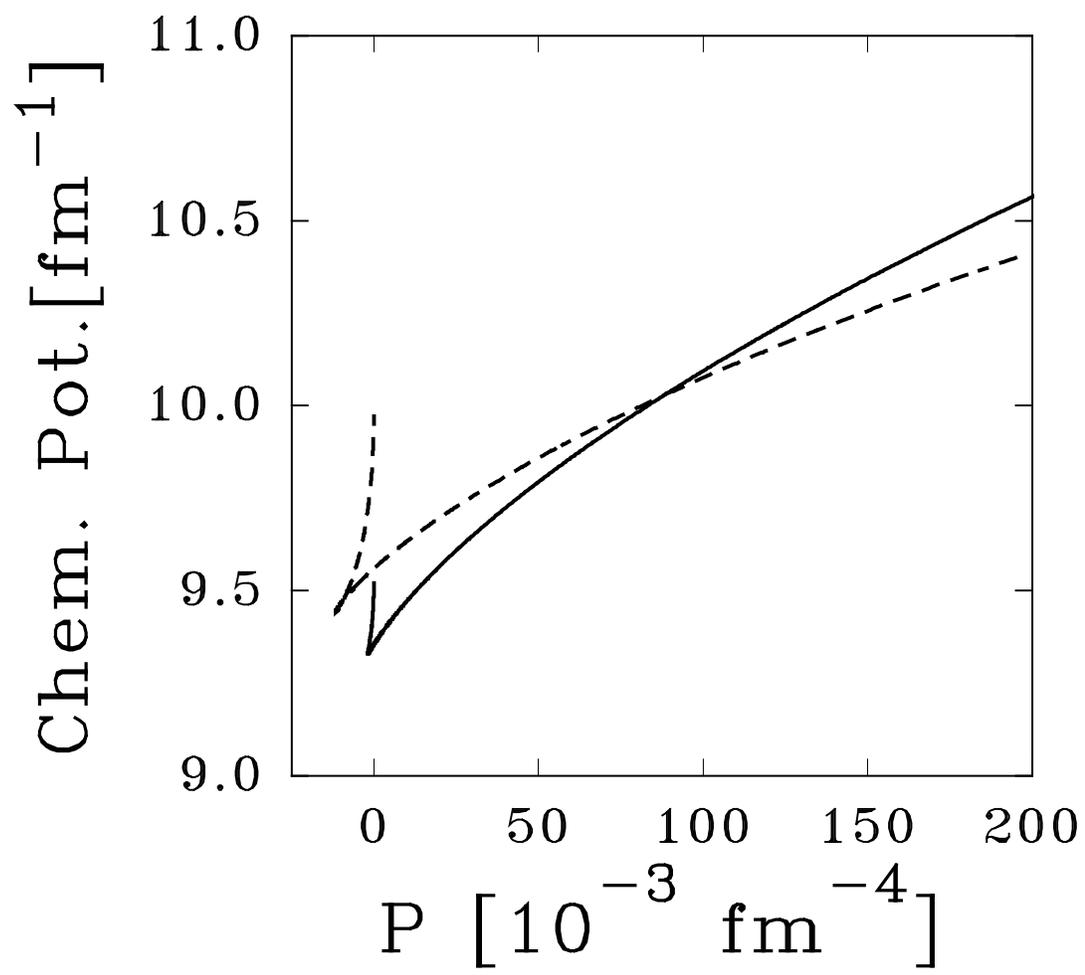
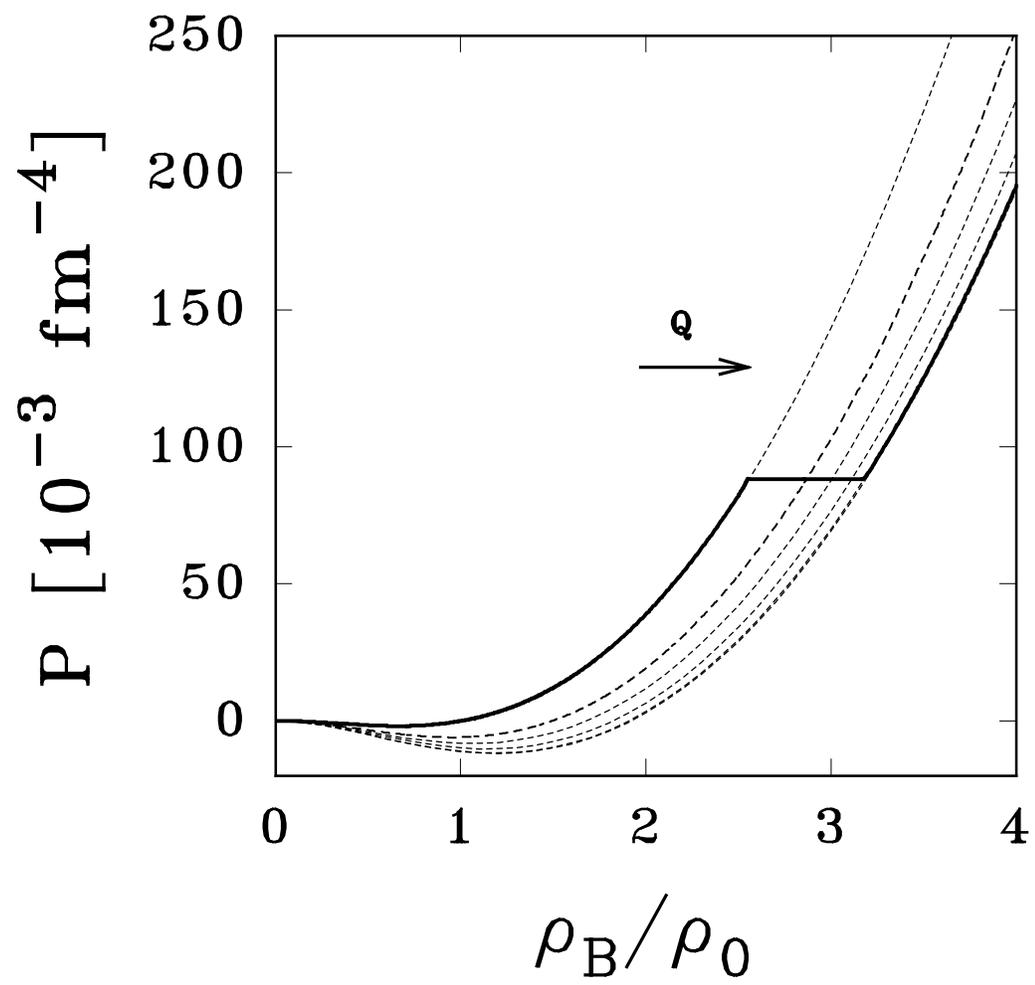
\end{document}